\documentclass[amssymb,aps,prl,twocolumn,groupedaddress]{revtex4-1}
\usepackage{graphics,epsfig,amsmath}

\newcommand{\beq}{\begin{eqnarray}}
\newcommand{\eeq}{\end{eqnarray}}
\newcommand{\p}{\partial}

\newcommand{\hs}[1]{\hspace{#1 mm}}
\newcommand{\bpm}{\begin{pmatrix}}
\newcommand{\epm}{\end{pmatrix}}

\newcommand{\ba}{\left(\begin{array}}
\newcommand{\ea}{\end{array} \right)}

\def\ee{\end{eqnarray}}

\newcommand{\nn}{\nonumber}
\def\p{\partial}

\def\=:{=\hspace{-.7em}\raisebox{1.1ex}{.}\hspace{.1em}\raisebox{-0.2ex}{.} }

\def\nn{\nonumber\\}
\begin{document}

\title{ 
Conformal symmetry of trapped Bose-Einstein condensates\\ and massive Nambu-Goldstone modes
}

\author{Keisuke~Ohashi}
\email[]{keisuke084@gmail.com}
\author{Toshiaki~Fujimori}
\email[]{toshiaki.fujimori018@gmail.com}
\author{Muneto~Nitta}
\email[]{nitta@phys-h.keio.ac.jp}

%\thanks{}
\affiliation{Department of Physics \& Research and Education Center 
for Natural Sciences, Keio University,
Hiyoshi 4-1-1, Yokohama, Kanagawa 223-8521, Japan}

\date{\today}

\begin{abstract}
The Gross-Pitaevskii or 
nonlinear Schr\"odinger equation relevant to 
ultracold atomic gaseous Bose-Einstein condensates 
possesses a modified Schr\"odinger symmetry 
in two spatial dimensions,  
in the presence of a harmonic trapping potential, 
an (artificial) constant magnetic field (or rotation), 
and an (artificial) electric field of a quadratic electrostatic potential.  
We find that a variance and a center of a trapped gas 
with or without a vorticity can be regarded 
as massive Nambu-Goldstone (NG) modes 
associated with spontaneous breaking of 
the modified Schr\"odinger symmetry. 
We show that the Noether theorem for the modified Schr\"odinger symmetry gives universal equations of motion 
which describe exact time evolutions of the trapped gases such as  
a harmonic oscillation,  a cyclotron motion, and  a breathing  oscillation with frequencies determined 
by the symmetry independent of the details of the system. 
We further construct an exact effective action for all the NG modes.
\end{abstract}

%\keywords{}

\maketitle
\section{Introduction}

Bose-Einstein condensations (BEC) were realized in
dilute ultracold atomic Bose gases in 
a trapping potential \cite{Anderson:1995,Davis:1995,Bradley:1995}.
The ideal feature of this system is 
its unprecedented controllability 
such as temperature, number of atoms, 
and strength and sign of interactions. 
BECs can be accurately described 
by the Gross-Pitaevskii (GP) or nonlinear Schr\"odinger equation 
\cite{Pitaevskii:2003}. 
BEC offers an ideal system with 
spontaneous symmetry breakings.  
In particular, spinor BECs have rich patterns of symmetry breakings 
and rich topology of order parameter spaces 
allowing various topological excitations 
\cite{Kawaguchi:2012ii}.

The symmetry of the free Schr\"odinger equation 
is known as the Schr\"odinger symmetry 
(or nonrelativistic conformal symmetry)
\cite{Hagen:1972pd,Niederer:1972zz}, 
consisting of the Galilean symmetry, the dilatation and 
the special Schr\"odinger transformation.  
In the presence of generic nonlinear interactions, 
the Schr\"odinger symmetry is explicitly broken to 
the inhomogeneous Galilean symmetry 
except in two spatial dimensions, 
where it remains intact if the non-linear term is limited to
a quartic (two-body) interaction term relevant for BECs, as long as the quantum anomaly is ignored. (See  \cite{2010PhRvL.105i5302O,Hu:2011} for recent discussions.)
 The presence of a trapping potential 
is necessary to realize BEC experimentally but 
it would break translational symmetry explicitly,
so that the Schr\"odinger symmetry in two spatial dimensions 
(Galilean symmetry in other dimensions) 
would be an approximate symmetry. 

In this paper, we show that 
in the presence of a harmonic trap, 
the Schr\"odinger (Galilean) symmetry 
including translational symmetry 
is not explicitly broken but is just modified 
as in the system of the $2d$ harmonic oscillator \cite{Niederer:1973tz}.  
This is an extension and completion of the previous study 
on the $O(2,1)$ subgroup \cite{Pitaevskii:1997,Ghosh:2001an} and 
the Galilean subgroup \cite{Ripoll:2001}. 
The Schr\"odinger symmetry in a harmonic trap 
was also discussed in the context of 
non-relativistic conformal field theories
without account for the modified Schr\"odinger symmetry 
\cite{Nishida:2007pj,Doroud:2015fsz}.
The point is that a harmonic potential 
has the same form as one of the generators 
of the Schr\"odinger symmetry, 
which can be regarded 
as a chemical potential term.
We further show that a synthetic constant magnetic field (or rotation) and 
a synthetic electric field of a quadratic electrostatic potential 
further modify the Schr\"odinger symmetry. 
One of the peculiar features of the modified Schr\"odinger symmetry 
is that some of the generators explicitly depend on time.
%%%%
The following question now arises: what happens 
when such a modified symmetry is spontaneously broken?
Here, we study Nambu-Goldstone (NG) modes arising due to the broken  modified Schr\"odinger symmetry 
and show that they are so-called massive NG modes 
\cite{Nicolis:2012vf,Nicolis:2013sga,Watanabe:2013uya,
 Takahashi:2014vua} resulting from the generators depending on time. 
Our physical interpretation is 
different from previous studies where authors interpreted, 
without knowing a modified symmetry, 
that the NG modes become massive 
because the original symmetry is explicitly broken 
by the chemical potential. 
A typical feature of massive NG modes is that
their excitations oscillate in time. 
As was done for the $O(2,1)$ subgroup 
\cite{Pitaevskii:1997,Ghosh:2001an} and 
the Galilean subgroup \cite{Ripoll:2001}, 
we further apply the full modified Schr\"odinger symmetry 
to collective motion of a trapped gas with or without a vorticity
and find their  exact time evolutions:
the trapped gas moves as a charged particle 
in the   artificially introduced external magnetic and electric fields,
and at the same time it breathes. 
The equations of motion describing 
such time evolutions are determined 
by the modified Schr\"odinger symmetry 
and hence they are universal equations 
which are independent of the details such as 
states of the gas and the strength of the nonlinear interaction. 
We also construct an exact effective action for massive NG modes. 
Our work will provide an example of 
massive NG modes in a physical system 
which can be tested experimentally in a laboratory.

%%%%%%%%%%%%%%%%%%%%%%%%%%%%%%%%%%%%%%%%%%
\section{Modified Schr\"odinger Symmetry}
We consider the nonlinear Schr\"odinger system 
in $d$ spatial dimensions described by the action 
\beq
S = \int dt \, d^{d}x \left\{  i \bar \phi \! \stackrel{\leftrightarrow}{{\cal D}}_t \! \phi
- \frac{1}{2m} |{\cal D}_i \phi|^2 - \frac{\kappa}{2} |\phi|^4 \right\}, 
\label{eq:action}
\eeq
where the scalar field $\phi$ is coupled to an external gauge field 
through the covariant derivative 
${\cal D}_M \phi=(\partial_M -i A_M) \phi$ 
with $M=t,1,\cdots,d$.
In the case of the free Schr\"odinger system ($\kappa = A_M = 0$), 
the action has the so-called {\it Schr\"odinger symmetry} 
or nonrelativistic conformal symmetry. 
To write down the explicit forms of the symmetry transformations, 
it is convenient to introduce the following 
set of time-independent Hermitian operators $\{ \underline{\mathbf Q_a} \}$ 
acting on $\phi$: 
\begin{eqnarray}
&\underline{{\bf Q}_t} \equiv i \partial_t, \quad
\underline{{\bf P}_i} \equiv -i \partial_i, \quad 
\underline{{\bf M}_{ij}} \equiv  i(x^j \partial_i-x^i\partial_j), \quad& 
\nonumber \\
&\underline{{\bf B}_i} \equiv m x^i,\quad
\underline{{\bf D}} \equiv i x^i\partial_i + i \frac{d}2, \quad 
\underline{{\bf C}} \equiv \frac{1}{2} m (x^i)^2, \quad&
\end{eqnarray}
and $\underline{{\bf N}} \phi \equiv \phi$. 
Then the generators $\{ \mathbf Q_{a} \}$ of 
the infinitesimal Schr\"odinger symmetry
$\delta \phi = \epsilon^{a} \mathbf Q_{a} \phi$
can be written as the following linear combinations of $\{ \underline{\mathbf Q_a} \}$: 
\begin{eqnarray}
{\bf Q}_t &=& \underline{{\bf Q}_t} ~~~~~~~~~~~~~~~~~~~ (\mbox{time translation}), \nonumber \\
{\bf N} &=&  \underline{{\bf N}}  ~~~~~~~~~~~~~~~~~~~\mbox{ (phase rotation)}, \nonumber \\
{\bf P}_i &=& \underline{{\bf P}_i} ~~~~~~~~~~~~~~~~~~ \mbox{ (spatial translation)}, \nonumber \\
{\bf M}_{ij} &=& \underline{{\bf M}_{ij}}~~~~~~~~~~~~~~~~~  \mbox{(spatial rotation)}, \nonumber \\
{\bf B}_i &=& \underline{{\bf B}_i} - t \underline{{\bf P}_i} ~~~~~~~~~~~  \mbox{(Galilean boost)}, \nonumber \\
{\bf D} \, &=& \underline{{\bf D}} + 2 t \underline{{\bf Q}_t} ~~~~~~~~~~ \mbox{(dilatation)}, \nonumber \\
{\bf C} \, &=& \underline{{\bf C}} + t \underline{{\bf D}} + t^2 \underline{{\bf Q}_t} ~~~
\mbox{(special Schr\"odinger)}.
\quad 
\end{eqnarray}
These operators form the Schr\"odinger algebra 
\cite{Hagen:1972pd,Niederer:1972zz}, 
whose nontrivial commutation relations are given by
\begin{eqnarray}
&[{\bf Q}_t, {\bf B}_i]=-i{\bf P}_i, ~~ [{\bf C},{\bf P}_i]=i {\bf B}_i, ~~~[{\bf B}_i,{\bf P}_j]=i m \, \delta_{ij}{\bf N}, & \nonumber \\
&{}[{\bf Q}_t,{\bf C}]=i{\bf D}, ~~~~~~~~
%&{}[{\bf M}_{ij}, {\bf P}_k]=i({\bf P}_j\delta_{ik}-{\bf P}_i\delta_{jk}), & \nonumber\\
%&{}[{\bf M}_{ij}, {\bf B}_k]=i({\bf B}_j\delta_{ik}-{\bf B}_i\delta_{jk}), & \nonumber\\
[{\bf D}, {\bf Q}_{a}]=i \Delta_{a} {\bf Q}_{a}, &
\end{eqnarray}
and $\{{\bf M}_{ij}\}$ forms the $\mathfrak{so}(d)$ algebra under which ${\bf P}_i$ and ${\bf B}_i$ transform as vectors. 
The constants $\Delta_a$ are conformal weights, given by 
$\Delta_{t}=-2$, $\Delta_{{\bf P}}=-1$,
 $\Delta_{{\bf M}}=\Delta_{{\bf N}} = \Delta_{{\bf D}}=0$,  
$\Delta_{{\bf B}}=1$ and $\Delta_{{\bf C}}=2$. 

Although the Schr\"odinger symmetry is explicitly broken 
once a generic external gauge field $A_M$ is turned on, 
the quadratic part of action (\ref{eq:action}) is invariant 
under a \textit{modified Schr\"odinger symmetry} 
for some specific external fields $A_M$. 
In this paper, we consider the external fields of the form
\begin{eqnarray}
A_i(x)  &=& m (a_i-b x^i-\omega_{ij}x^j), \\
A_t(x)  &=& \alpha + e_i x^i +\frac12 g_{ij} x^ix^j,
\label{eq:EleMagPotential}
\end{eqnarray}
where summation over the repeated indices is understood.
Here,  $b$, $\alpha$, $a_i$ and $e_i$, $\omega_{ij}$ 
(antisymmetric) 
are all constants in $x^i$ but can depend on time. The matrix $g_{ij}$ is  given by 
$g_{ij} = m \omega_{ik} \omega_{jk} - (\lambda-mb^2) \delta_{ij}$ with a positive definite parameter $\lambda$. 
For this choice of the external gauge field, 
the equation of motion of the action (\ref{eq:action}) 
reads
\begin{eqnarray}
i \partial_t  \phi &=& \left( -\frac{1}{2m} \partial_i^2 - \mu^{\hat a} \underline{{\bf Q}_{\hat a}} + \kappa |\phi|^2 \right) \phi, 
\label{eq:modScheq}
\end{eqnarray}
where 
$\mu^{\hat a} \underline{{\bf Q}_{\hat a}}$ are the generalized chemical potential terms 
\begin{eqnarray}
\mu^{\hat a} \underline{{\bf Q}_{\hat a}} = 
\mu \underline{\bf N}
+ a^i \underline{\bf P}_i + b \, \underline{\bf D} 
+ \frac{\omega_{ij}}{2} \underline{\bf M}_{ij}
+ c^i \underline{\bf B}_i - \frac{\lambda}{m} \underline{\bf C} ~~
\label{eq:addterms}
\end{eqnarray}
with $c_i = e_i / m + b a_i - \omega_{ij} a^j$ and $\mu = \alpha - \frac{1}{2} m a_i^2$.
Here $\hat a$ is the label of the operators 
excluding the time translation $\underline{\mathbf Q_t}$.
Note that  $\mu$ is the standard chemical potential and $\lambda$ is the strength of the harmonic trap.
Here we regard $\mu^{\hat a}$ as controllable parameters and 
allow them to have time dependence, $\mu^{\hat a}=\mu^{\hat a}(t)$.  

Let us look for the symmetry of the quadratic part of the action (\ref{eq:action}) 
with non-vanishing $\mu^{\hat a}$. 
We assume that the infinitesimal transformation of $\phi$ takes the form
$\delta \phi = [ \xi^M(t,x^i) \partial_M +\Xi(t,x^i) ] \phi$. 
By requiring that $\delta \phi$ is a symmetry of the quadratic part of the action
or,  equivalently, the corresponding equation of motion
\beq
i \p_t \phi = \mathbf H_L \phi, \hs{5} 
\mathbf H_L \equiv - \frac{1}{2m} \p_i^2 - \mu^{\hat a} \underline{\mathbf Q}_{\hat a}, 
\eeq
we can show that the infinitesimal transformation takes the form
$\delta \phi = - i \eta^{a}(t) \underline{{\bf Q}}_{a} \phi.$
The time dependence of the coefficients $\eta^t(t)$ and $\eta^{\hat a}(t)$ 
can be determined by solving the following equations
\begin{eqnarray}
\dot {\eta}^t(t) = \eta^{\hat a}(t) {\cal I}_{\hat a} ,\quad
\dot {\eta}^{\hat a}(t) = \eta^{\hat b}(t) {\cal J}_{\hat b}{}^{\hat a}+\eta^t(t) \dot {\mu}^{\hat a}(t), 
\label{eq:ParaEvolution}
\end{eqnarray}
where $\dot X(t)=dX(t)/dt$ and 
$({\cal I}_{\hat a}, {\cal J}_{\hat a}{}^{\hat b})$ are the coefficients in the commutation relation 
$[{\bf H}_{\rm L}, \underline{{\bf Q}_{\hat a}}] = i {\cal J}_{\hat a}{}^{\hat b} \underline{{\bf Q}_{\hat b}} + i {\cal I}_{\hat a} {\bf H}_{\rm L}$.

Since Eqs.\,(\ref{eq:ParaEvolution})  
are homogeneous linear differential equations, 
the general solution takes the form $\eta^a(t) = \epsilon^b \, \mathcal G_b{}^a(t)$,
with a time-dependent matrix $\mathcal G_b{}^a(t)$ and constant parameters $\epsilon^a$. 
Thus we find that the transformations generated by the operators
\begin{eqnarray}
{\bf Q}_a(t) \equiv \mathcal G_{a}{}^{b}(t) \underline{{\bf Q}_{b}},
\label{eq:Qh}
\end{eqnarray} 
are the symmetry of the quadratic part of the action.
If $\mu^{\hat a}$ are independent of $t$,  
the matrix $\mathcal G_b{}^a(t)$ can be written as $\mathcal G_{a}{}^{b}(t) = (e^{{\cal H} t})_{a}{}^{b}$,
by using the matrix $({\cal H})_{\hat a}{}^{\hat b} = {\cal J}_{\hat a}{}^{\hat b} $, 
$({\cal H})_{\hat a}{}^t = {\cal I}_{\hat a}$, and $({\cal H})_t{}^b=0$. 
These generators reduce to those of the Schr\"odinger group 
when the parameters $\mu^{\hat a}$ are turned off. 
Furthermore, we can show that 
$\{ {\bf Q}_a(t) \}$ has the same algebraic structure as the Schr\"odinger algebra. 
Therefore, in the presence of non-zero $\{\mu^a\}$, 
{\it the Schr\"odinger symmetry is not broken, but modified}. 
For instance, when $\mu^a=0$ except for the trapping potential $\lambda$, 
the generators of the translation and the Galilean boost oscillate: 
\begin{eqnarray}
{\bf P}_i(t) &=& \cos ( \omega_\lambda t ) \underline{{\bf P}_i} + \omega_\lambda \ \sin( \omega_\lambda t ) \underline{{\bf B}_i}, \nn
{\bf B}_i(t) &=& \cos (\omega_\lambda t ) \underline{{\bf B}_i} - \omega_\lambda^{-1} \! \sin( \omega_\lambda t ) \underline{{\bf P}_i}, 
\end{eqnarray}
with $\omega_\lambda=\sqrt{\lambda/m}$. 

Next let us turn on the nonlinear term $\kappa |\phi|^{2\beta }$ in the action.  
This term is invariant if  $\beta=1+2/d$ in $d$ spatial dimensions,
so that {\it the full modified Schr\"odinger group is the symmetry of the nonlinear action}. 
For $\beta \not =1+2/d$, it is explicitly broken to 
{\it the modified inhomogeneous Galilean symmetry}
which consists of ${\bf Q}_t, {\bf P}_i,{\bf M}_{ij},{\bf B}_i,{\bf N} $.
In the following, we focus on the $d=2, \beta=2$ case, 
where the action has the full modified Schr\"odinger symmetry. 

\section{Noether Charges}
Let us consider localized configurations in $d=2$ dimensions.
The scalar $\phi$ asymptotically behaves as $\phi \sim \exp(-\frac{1}{2} g |x^i|^2 + {\cal O}(x))$ with 
$g = \sqrt{m(\lambda-b^2 m)} - imb$, 
so that stable localized solutions exist when $\lambda-b^2m>0$. 
For such a localized object, the Noether charges $Q_a$ of 
the modified Schr\"odinger symmetry take finite values. 
It is convenient to rewrite the conserved charges $Q_a$ as 
\beq
Q_a = \mathcal G_{a}{}^{b}(t) \, \underline{Q_{b}}(t), 
\eeq
where $\underline Q_a$ are the following ``charges" defined 
by using the operators $\underline{\mathbf Q_{\hat a}}$ and ${\bf H}_L$: 
\begin{eqnarray}
\underline{Q_{\hat a}} \equiv \int d^2x \, \bar \phi \, \underline{{\bf Q}_{\hat a}} \phi, \quad 
\underline{Q_t} \equiv \int d^2x \left( \bar \phi \, {\bf H}_L \phi +\frac{\kappa}{2} |\phi|^4 \right). \quad \label{eq:def_charge} 
\end{eqnarray} 
Although $\underline{Q_a}$ are not conserved charges, 
their time dependence can be determined from the Noether theorem $\dot Q_a = 0$ as
$\underline{\dot Q_{a}} = - \left( \mathcal G^{-1} \p_t \mathcal G \right)_a{}^b \, \underline{Q_b}$, 
or equivalently  
\begin{eqnarray}
\underline{\dot Q_t} = -\dot \mu^{\hat a} \underline{Q_{\hat a}}, \hs{7} 
\underline{\dot Q_{\hat a}} = - {\cal J}_{\hat a}{}^{\hat b} \underline{Q_{\hat b}} - {\cal I}_{\hat a} \underline{Q_t} . 
\label{eq:chargeEq} 
\end{eqnarray}
The physical meaning of this Noether's theorem can be understood as follows.
The first equation in Eq.\,\eqref{eq:chargeEq} describes 
the relation between the energy $E=\underline{Q_t}$ 
and the time dependence of the parameters $\mu^{\hat a}(t)$.
Since the phase rotation is the symmetry of the modified action, 
one of the second equations in Eq.~\eqref{eq:chargeEq} gives the conservation law $\underline{\dot N}=0$ 
for the particle number $N = \underline{N} = \int d^2 x |\phi|^2$.

Defining the center of the gas by the first moment of the charge density
\begin{eqnarray}
\underline x^i(t) \equiv \frac{1}{m} \frac{\underline{B}_i }{N\,} = \frac{1}{N} \int d^2 x \, x^i |\phi|^2 ,
\end{eqnarray}
and the momentum
\beq
\underline{p_i}(t) \equiv \frac{\underline{P}_i}{N} = - \frac{i}{N} \int d^2 x \, \bar \phi  \,\p_i \phi, 
\eeq
we can interpret the equation for $\underline{\dot B_i}$ 
as the relation between the velocity and the momentum
in the external magnetic field $m \dot {\underline x}^i = \, \underline{p_i} - A_i( \underline x )$.
Furthermore, eliminating the momentum $\underline{p_i}$ 
from its equation of motion 
$\underline{\dot p_i} = -\p_i[ ( \underline{p_i} - A_i)^2/2m -  A_t ]_{x^i = \underline{x}^i}$,
we find that the gas behaves as a charged particle 
\begin{eqnarray}
m\, \ddot {\underline x}^i = \dot {\underline x}^j F_{ij}(\underline x) + F_{i0}(\underline x),
\end{eqnarray}
in the synthetic magnetic field 
$F_{ij} \equiv \p_i A_j - \p_j A_i = 2m \omega_{ij}$ 
and the synthetic  electric field $F_{i0} \equiv \p_i A_t - \p_t A_i$ 
for the gauge potentials in Eq.~(\ref{eq:EleMagPotential}). 
This equation can be regarded as a version of Ehrenfest's theorem generalized so that it works even
 with the presence of the nonlinear term, $\kappa\not=0$.  
 Since this property comes from the modified inhomogeneous Galilean symmetry,  it
 holds for any interaction in any dimension \cite{Kohn:1961}.  
Its dynamics is described by a combination of the cyclotron motion caused by the magnetic field with a frequency $\omega \equiv \omega_{12}$ and the harmonic oscillation due to the trapping potential with a frequency 
\begin{eqnarray}
\omega_\lambda \equiv \sqrt{\frac{\lambda}m-b^2} \label{eq:cyclotron}.
\end{eqnarray}
Eventually the frequencies of the combined motions are 
$\pm(\omega\pm \omega_\lambda)$. 

The equation of motion for the angular momentum $\underline M_{12} = - i \int d^2 x \, \bar \phi (x_1 \p_2 - x_2 \p_1) \phi $ 
can be rewritten as the conservation law for the internal angular momentum,
$\dot S = 0$ with $S \equiv \underline{M}_{12}/N - (\underline{x_1} \, \underline{p_2} - \underline{x_2} \, \underline{p_1} )$. 

The charge $\underline D(t) = \frac{i}{2} \int d^2 x \, \bar \phi \{x^i, \p_i \} \phi$ satisfies
\beq
\underline{\dot D} &=& 2 E + \sum_{\hat a} (2 - \Delta_{\hat a}) \mu^{\hat a} \underline {Q_{\hat a}}. 
\label{eq:eqD}
\eeq
Defining $\sigma(t)$ by using the second moment (variance)
\begin{eqnarray}
\sigma(t) 
\equiv \frac{2}{m} \frac{\underline C}{N} - (\underline x^i)^2 
= \frac{1}{N} \int d^2x \, (x^i - \underline x^i)^2 |\phi|^2, 
\end{eqnarray}
we find that $\underline \sigma$ satisfies 
\beq
\dot \sigma &=& 2 b \sigma - \frac{2}{m} \left( \frac{\underline D}{N} 
+\underline x_1 \underline{p_1} + \underline x_2 \underline{p_2} \right). 
\eeq
Eliminating $\underline D$, 
we find that $\sigma$ satisfies the equation %for a harmonic oscillator
\begin{eqnarray}
m \ddot {\sigma} &=& - 2 \left( 2 \lambda - 2 m b^2 + m \dot b \right) \sigma \nn
&-& 4 \left[ \frac{m}{2} \left( \underline{\dot x}^i \right)^2 - A_t(\underline x) - \omega_{12} S - \frac{E}{N} \right] .\quad 
\end{eqnarray}
When all $\mu^a$ are constant in time, 
this reduces to a harmonic oscillator with  frequency 
$\pm 2\omega_\lambda$. 

As was pointed out in \cite{Pitaevskii:1997,Kagan:1996}, 
the frequencies   
$\pm(\omega \pm \omega_\lambda)$ 
for $(\underline x_i, \underline p_i)$ and 
$\pm 2 \omega_\lambda$ for $(\sigma,\underline D)$ 
are universal constants determined by the symmetry. 
They are related to the conformal weights 
and the angular momenta of primary operators
via the state-operator correspondence \cite{Nishida:2007pj}. 
In non-relativistic conformal field theories, 
the conformal weight $\Delta_{\mathcal O}$ 
and the angular momentum $j_{\mathcal O}$ of a primary operator $\mathcal O$, 
i.e.,  an operator satisfying 
$[\mathbf B_i, \mathcal O] = [\mathbf C, \mathcal O]  = 0$, 
is related to the energy eigenvalue $E_{\mathcal O}$ of the Hamiltonian with the harmonic potential as 
$E_{\mathcal O} = \Delta_{\mathcal O} \omega_\lambda + j_{\mathcal O} \omega $.
This implies that the state-operator correspondence 
maps the primary operators 
$\mathbf B_\pm \equiv \mathbf B_1 \pm i \mathbf B_2$ 
and $\mathbf C$ to the excited states of the gas with energy 
$E_{\mathbf B_\pm} = \omega_\lambda \pm \omega$ and 
$E_{\mathbf C} = 2 \omega_\lambda$, 
which correspond to the first excited states generated 
by the oscillation modes $(\underline x_i, \underline p_i)$ 
and $(\sigma,\underline D)$, respectively.

The charges $\underline{Q_a}$ mentioned above 
will turn out to be identified with NG modes 
generated by their ``conjugate" generators, as we will see later. 
%When $\mu^a$ and $b$ are constant the above equation is for a harmonic oscillator,

%%%%%%%%%%%%%%%%%%%%%%%%%%%%%

%%%%
\section{Massive Nambu-Goldstone modes and exact 
time evolution}
We have seen above that the modified Sch\"odinger symmetry 
gives universal information, even when there are vortex-like or wall-like objects.  
Here, 
 we focus on static axially symmetric configurations $\phi=\phi^{\rm sol}$ 
 as typical examples,
 setting $a^i = b = c^i = \dot \omega = \dot \mu = \dot \lambda = 0$ for simplicity. 
 We  define the vorticity $l\in \mathbb Z$ as the winding number of the phase of $\phi$ around the center of the trap.   
The solution is invariant under ${\bf M}_{12} - l {\bf N}$ and thus 
the $\omega$ dependence of the solution appears only in the form $\tilde \mu \equiv \mu + l \omega$. 
 Some numerical solutions and $\tilde \mu $ dependences  of the charge $N$ and the energy $E$
 are shown in the appendix. 

The axially symmetric static solutions break
the modified Schr\"odinger symmetry to its subgroup
generated by ${\bf Q}_t = i\partial_t$ and ${\bf M}_{12} - l {\bf N}$,
so that there must appear NG modes around the solution $\phi=\phi^{\rm sol}$. 
Such  NG modes are generated by the symmetry
whose transformation parameters are promoted to dynamical degrees of freedom
$\epsilon^a \rightarrow \epsilon^a(t)$, 
\beq
\delta \phi = \sum_a \epsilon^a(t) \mathbf Q_a(t) \phi^{\rm sol} 
= \sum_a \delta \pi^a(t) \underline{\mathbf Q_{a}} \phi^{\rm sol},
\eeq
where we have redefined the degrees of freedom 
by absorbing the time dependence of the generator 
$\mathbf Q_a(t) = \mathcal G_a{}^b(t) \underline{\mathbf Q_a}$ 
as $\delta \pi^a(t) \equiv \epsilon^a(t) \mathcal G_a{}^b(t)$. 
The important difference from the case of the standard symmetry breaking is that 
some of $\delta \pi^a$ are massive NG modes. 
%discussed in the literature 
%\cite{Nicolis:2012vf,Nicolis:2013sga,Watanabe:2013uya, Takahashi:2014vua}.

%As discussed in \cite{Watanabe:2014pea}, 
The kinetic terms of the massive NG modes 
in the low-energy effective action are 
determined by the commutators  of the corresponding generators evaluated for the solution. 
The nontrivial commutation relations are given by
 \begin{eqnarray}
\int d^2x \, \bar \phi^{\rm sol} [\underline{\bf B}_i, \underline{\bf P}_j]  \phi^{\rm sol} &=& \delta_{ij} m \, N, 
\nonumber \\
\int d^2x \, \bar \phi^{\rm sol} [\underline{\bf D}, \underline{\bf C}] \phi^{\rm sol} &=& 2 \underline C.
\label{eq:conjugation}
\end{eqnarray}   
This implies that the NG modes for $\{ \underline{\mathbf P_i}, \underline{\mathbf B_i} \}$
and $\{ \underline{\bf C}, \underline{\bf D}\}$ form canonical conjugate pairs. 
The number of generators in the former (latter) pair is four (two), 
while the number of corresponding NG modes is two (one), which is a half of the number of 
generators, 
since the (non)commutation relations in Eq.\,(\ref{eq:conjugation}) give rise to type-B NG modes 
\cite{Watanabe:2012hr,Hidaka:2012ym}. 

On the other hand, since the generator ${\bf N}$ commutes with the other operators, giving rise to a type-A NG mode,
a certain conjugate degree of freedom $\nu(t)$ must be introduced 
to construct a kinetic term for the NG mode (phonon) $\theta(t)$.
Taking into account the fact that $\theta(t)$ appears in the effective Lagrangian
so that $\p L_{\rm eff} / \p \dot \theta = N(\tilde \mu)$, 
it is natural to introduce the conjugate mode $\nu(t)$ 
by shifting  $\tilde \mu$ dependence as $\tilde \mu+\nu$.    

The low-energy effective action describing the dynamics of the NG modes $\pi^a(t)$
can be determined from the time dependent ansatz
obtained by acting the finite transformation $U$ on the static solution $\phi^{\rm sol}$ as 
$\phi(t) = U(\pi^a(t)) \, \phi^{\rm sol}$.
Since $\{ \underline{\mathbf Q_a} \}$ is the basis of $\{{\bf Q}_a(t)\}$, 
the operator $U$ can be written as 
\begin{eqnarray}
U \! \! &=& \! \! \exp \left( \nu \p_\mu \right)  \exp \left( -i \underline{x}^i \underline{\mathbf  P}_i \right) \nn 
&\times& \! \! \exp \left( i \underline{p_i} / m \, \underline{\mathbf B_i} - i \chi \underline{\bf C} \right)
\exp \left(- i\theta {\bf N} - i \rho \underline{\bf D} \right),
\end{eqnarray}
where we have introduced the shift operator for $\mu$ 
to include the conjugate mode $\nu$ as we explained.
Explicitly, $\phi$ can be written as  
\begin{eqnarray}
\phi = A \, \phi^{\rm sol} \big( e^{\rho} (x^i - \underline x^i) , \, \tilde \mu + \nu \big),
\label{eq:boostsol}
\end{eqnarray}
with $A \!=\!\exp \left[ \rho - i \theta + i \, (x - \underline x)^i \underline{p_i} -i m |x - \underline x|^2 \chi/2 \right]$.
The dynamical degrees of freedom of the NG modes 
$\pi^a(t) = \underline x^i, \underline{p_i}, \rho, \chi$ and $\nu(t)$
are essentially equivalent to those of the charges $\underline{Q_{\hat a}}$. 
This fact can be seen by from the relation between $\pi^a(t)$ and $\underline{Q_a}(t)$ 
obtained by substituting the ansatz into the charges $\underline{Q_a}$ in Eq.\eqref{eq:def_charge}, 
\begin{eqnarray}
\underline N &=& N^{\rm sol}, \quad
\underline{B_i} = m \underline x^i N^{\rm sol}, \quad 
\underline{P_i} = \underline{p_i} N^{\rm sol} \nn
\underline C &=& e^{-2\rho} \underline C^{\rm sol} + \frac{m}{2} (\underline x^i)^2 N^{\rm sol}, \nn
\underline D &=& 2 \chi e^{-2\rho} \underline C^{\rm sol} - \underline x^i \underline{p_i} N^{\rm sol}, 
\end{eqnarray}
where $N^{\rm sol} =N^{\rm sol}(\tilde \mu + \nu)$ and 
$\underline C^{\rm sol} = \underline C^{\rm sol}(\tilde \mu+\nu)$ are 
the charges evaluated for the solution with the shifted chemical potential $\tilde \mu+\nu$. 
We can show that the time dependent ansatz $\phi(t)$ exactly satisfies the original equation of motion
when $\pi^a(t)$ have time dependence 
so that the equation of motion (\ref{eq:chargeEq}) for the charges $\underline{Q_a}$ 
and the equation for $\theta$, 
\begin{eqnarray}
\dot \theta = - \tilde \mu + (\tilde \mu+\nu) e^{2\rho} - \frac{\underline{p_i}^2}{2m} + \frac{\lambda}{2} \underline{x_i}^2,
\label{eq:eom_theta}
\end{eqnarray}
are satisfied.

By substituting the ansatz (\ref{eq:boostsol}) into the action (\ref{eq:action}),
we obtain the effective action $S_{\rm eff} = \int dt {\cal L}_{\rm eff}$
for the NG modes and $\nu(t)$ as,
\begin{eqnarray}
{\cal L}_{\rm eff} \! \! &=& \! \! N^{\rm sol} \left\{ \tilde \mu + \dot \theta + \underline{p_i} \underline{\dot x}^i - \frac{\underline{p_i}^2}{2m} - \frac{\lambda}{2} \underline{x_i}^2 -\omega_{ij} \underline p^i \underline x^j \right\}\quad \nn
&+& \! \! C^{\rm sol} \left\{ \dot \chi e^{-2\rho } -\chi^2 e^{-2\rho} - \frac{2\lambda}{m} \cosh(2\rho) \right\}.
\label{eq:L_effective}
\end{eqnarray}
We can easily obtain time evolutions of $\pi^a(t)$
by solving the equations of motion derived from this effective action. 
For instance, $\delta S_{\rm eff}/\delta \theta=0$ gives $\dot \nu(t)=0$ 
and $\delta S_{\rm eff}/\delta \nu=0$ gives the equation of motion 
\eqref{eq:eom_theta} for the massless mode $\theta$ 
through the following identity for a static solution:
\begin{eqnarray}
0 = \left[\frac{2\lambda}{m} \frac{\partial}{\partial \mu} C^{\rm sol} -\mu \frac{\partial}{\partial \mu} N^{\rm sol} \right]_{\mu \to \tilde \mu+\nu}, 
\label{eq:NCrelation2}
\end{eqnarray}
which can be derived from the Noether theorem in 
Eq.\,(\ref{eq:chargeEq}) as shown in the appendix.
Thus one can easily construct time-dependent solutions 
from a static solution $\phi^{\rm sol}(x, \tilde \mu+\nu)$. 
Animations of a time-dependent solution with $m=\kappa=\omega=\mu=l=1$ and $ \lambda=1/2$ are shown in the ancillary file.

\section{Summary}

We have found the modified Schr\"odinger symmetry 
of the GP equation in two spatial dimensions,  
in the presence of a harmonic potential, 
(artificial) constant magnetic field (or rotation) 
and an (artificial) electric field of a quadratic electrostatic potential. 
We have studied NG modes of 
the spontaneously broken modified Schr\"odinger symmetry, 
and found that 
spontaneous breaking of $\underline{\bf B}_i$ and $\underline{\bf P}_i$ 
($\underline{\bf D}$ and $\underline{\bf C}$), 
which do not commute as in Eq.\,(\ref{eq:conjugation}),  
 give rise to two (one) massive NG modes of type B,
 while that of ${\bf N}$  gives 
 rise to one type-A NG mode (phonon).
We have found that a variance in addition to a center of a gas can be
 regarded as massive NG modes 
 and derived the universal equations of motion from the Noether theorem.  
% that they must obey certain exact time-evolutions
% described by the equations of motion which do not depend on 
% the details such as states of the gas and 
% the coefficients of the non-linear term $\kappa$. 
We have further constructed the general boosted solution 
in which the trapped gas does a cyclotron motion with  frequency $\omega_{12}$ and 
a harmonic oscillation with  frequency $\omega_{\lambda}$ in 
Eq.~(\ref{eq:cyclotron}) and 
at the same time breathes with  frequency 
$2\omega_{\lambda}$.
Finally, we have constructed the exact effective 
action for  all the NG modes.
The frequencies 
%cyclotron frequencies in Eq.\,(\ref{eq:cyclotron}) 
%together with $omega_{12}$
%and the breathing frequency in Eq.\,(\ref{eq:freq}) 
of collective motions of BECs we have found 
should be observed in experiments in 
ultracold atomic gases.
The effective Lagrangian Eq.~\eqref{eq:L_effective} 
exactly describes the time evolution of the gas 
even when the parameters $\mu^a$ depend on time. 
It would be interesting to see if 
it is possible to adjust the time dependence of $\mu^a$ 
to find some characteristic time evolutions such as resonances and amplifications of the oscillations.

The universal frequencies $\pm(\omega \pm \omega_\lambda)$ 
and $\pm 2\omega_\lambda$ in Eq.~(\ref{eq:cyclotron}) 
should be observed in experiments in ultracold atom gases, 
as was done for the breathing mode $\pm 2 \omega_\lambda$ 
in the case of $a_i=b=\omega_{ij}=c_i=0$ in Eq.~\eqref{eq:addterms} \cite{Chevy:2002}.
The stability and robustness of massive NG modes against temperature effect and/or quantum fluctuations is one of the important issues which should be clarified.  In particular,  a quantum anomaly for the dilatation symmetry can generally arise and shift the frequencies  as was discussed in \cite{2010PhRvL.105i5302O,Hu:2011}. It would be interesting to study how that anomaly appears in the low-energy effective action.

If we focus only on the Galilean subgroup, 
we can extend our analysis to anisotropic harmonic potential 
in any dimensions, as was done for $d=2$ \cite{Ripoll:2001}. 

Although we have used numerical results as the static solutions $\phi=\phi^{\rm sol}$ in this work,
it would be interesting to discuss  them in fully analytic ways by taking some appropriate limits,  as is done in \cite{Biasi:2017pdp}.          

Our work should be extended to other NG modes in trapped BECs,
 such as a kelvin mode of a vortex line 
 \cite{Pitaevskii:1961,Donnelly, Takahashi:2015caa,Takahashi:2017ruq},  
a Tkachenko mode in a vortex lattice \cite{Baym:2003}, 
 and a ripple mode on a domain wall 
 (see, e.g.,~Refs.~\cite{Takeuchi:2013mwa,Takahashi:2015caa}).

The notion of modified symmetry should be applied to 
interpolation of relativistic and nonrelativistic NG modes, 
 studied
in a Lorentz invariant theory 
modified by a chemical potential term 
\cite{Kobayashi:2015pra} 
 (see also Refs.~\cite{Kobayashi:2014xua,Kobayashi:2014eqa}).

\begin{acknowledgments}
\subsection{Acknowledgments}
We thank Daisuke A. Takahashi for useful comments.
This work is supported by  the Ministry of Education,
Culture, Sports, Science (MEXT)-Supported Program for the Strategic
Research Foundation at Private Universities ``Topological Science''
(Grant No.~S1511006).
The work of M.~N.~is also supported in part by a Grant-in-Aid for
Scientific Research on Innovative Areas ``Topological Materials
Science'' (KAKENHI Grant No.~15H05855) from the MEXT of Japan, and 
by the Japan Society for the Promotion of Science
(JSPS) Grant-in-Aid for Scientific Research (KAKENHI Grant
No.~16H03984).
\end{acknowledgments}

\appendix
\section{Static Localized Solutions}
\label{sec:solution}
In this supplemental material, we discuss 
 static solutions in $d=2$ spatial dimensions. 
For simplicity, we set 
$a^i = b = c^i = \dot \omega = \dot \mu = \dot \lambda = 0$. 
Then the equation of motion for a static configuration takes the form
\beq
0 \! &=& \! \left( - \frac{1}{2m} \p_i^2 + \kappa |\phi|^2 - \omega \underline{\bf M}_{12} 
- \mu \underline{\bf N} + \frac{\lambda}{m} \underline{\bf  C} \right) \phi. ~~
\eeq

\begin{figure}[h]
 \begin{center}
   \includegraphics[width=7cm]{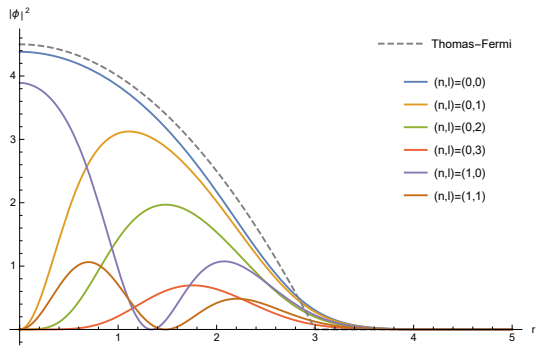}
 \end{center}
\caption{\footnotesize $|\phi|^2$ for axially symmetric solutions with $\kappa=\lambda=m=1$ and $\tilde \mu=4.5$. The horizontal axis is for the radial coordinate $r=\sqrt{|x^i|^2}$. The profiles oscillate below 
the dashed line, which gives the Thomas-Fermi approximation, and exponentially decay in the outer region with $r>3\, (=\sqrt{2\tilde \mu/\lambda})$. } 
\label{fig:DensityConfig}
\end{figure}

\begin{figure}[t]
 \begin{center}
   \includegraphics[width=7cm]{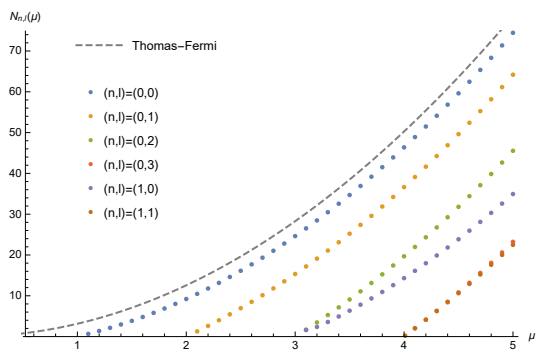}
 \end{center}
\caption{\footnotesize $N_{n,l} = N[\phi_{n,l}^{\rm sol}]$ with $\kappa=\lambda=m=1$ and $\omega=0$. 
Since $|\phi|$ becomes small in the limit $s \rightarrow s_c \equiv  2n + |l| + 1$, 
the field equation can be approximated as the linearized one, 
i.e. the Schr\"odinger equation for a harmonic oscillator. 
For $\kappa >0$, there is no solution below this limit, $\tilde \mu \le \sqrt{\lambda/m} \, s_c $.} 
\label{fig:Nmufig}
\end{figure}

\begin{figure}[t]
 \begin{center}
   \includegraphics[width=7cm]{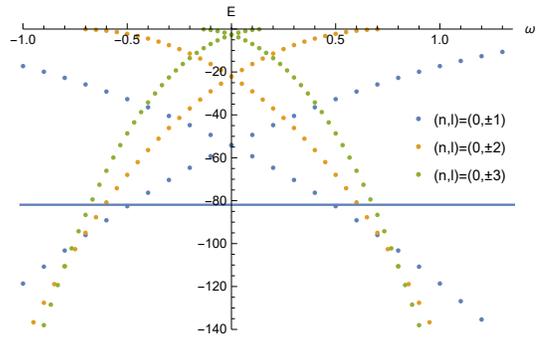}
 \end{center}
\caption{\footnotesize $E[\phi^{\rm sol}_{n,l}(\mu+n\omega)]$ 
with $\kappa=\lambda=m=1$ and $\mu=4.5$. 
The solid line corresponds to the solution with $(n,l)=(0,0)$, 
which has the lowest energy for small magnetic field $m \omega$.}
\label{fig:Energy}
\end{figure}

We assume the following axial symmetric solution, 
\begin{eqnarray}
\phi = \phi_{l,n}^{\rm sol} = (x_1 + i x_2)^l f_{n,l} (|x_i|),
\end{eqnarray}
where $f_{n,l}$ is a real smooth function and 
non-negative integers $n$ and $l$ are 
the number of nodes of $f_{n,l}$ and the winding number, respectively.
For each pair of integers $(n,l)$, 
there is a unique solution satisfying the boundary condition 
$\lim_{|x|\to \infty} f_{n,l}=0$ as shown in FIG.\ref{fig:DensityConfig}.
We can also show that the energy has the following parameter dependence,  with  $\tilde \mu \equiv \mu + l \omega$, 
\beq
E = \frac{\tilde \mu}{m \kappa} W_{n,l}(s), \hs{7} 
s \equiv \tilde \mu \sqrt{m/\lambda}, 
\eeq
where $W_{n,l}(s)$ is a function which depends on 
the dimensionless parameter $s$ and integers $n$ and $l$.
The $\mu$ dependence of $N$ and $\omega$ dependence of $E$ 
for the axially symmetric solutions are shown in FIG.\,\ref{fig:Nmufig} and FIG.\,\ref{fig:Energy}, respectively. 
%%%%%%%%%%%%%%%%%%%%%%%%%%%%%%%%%%%%%

%%%%%%%%%%%%%%%%%%%%%%%%%%%%%%%%%%%%%%

The Noether theorem in Eq.\,(15) in the letter for static solutions implies 
various relations between the charges $\underline{Q_a}$. 
For example, $\underline{P_i} = \underline{B_i} = 0$ for $\lambda \not = m \omega^2$, 
which means that the center of gas must be at the origin. 
For $\lambda = m \omega^2$, 
$\underline{B_i} = - \omega_{ij} \underline{P_j}$ can be arbitrary 
since the electric field $F_{i0}^{\rm}$ vanishes in this case. 
Furthermore, we find from Eq.\,(20) in the letter
that the total energy is related to $N$ and $\underline C$ as
\begin{eqnarray}
E = - \tilde \mu N + \frac{2\lambda}{m} \underline C.
\label{eq:ENC}
\end{eqnarray}
We can check this relation for the axially symmetric solutions
by using the fact that $N$ and $\underline C$ are given by
\begin{eqnarray}
N &=& - \frac{\partial E}{\partial \mu} = - \frac{1}{m \kappa} \left[ W_{n,l}(s) + s W_{n,l}'(s) \right], \\
\underline C &=& m \frac{\partial E}{\partial \lambda} = - \frac{1}{2} 
\frac{\tilde \mu s}{\kappa \lambda} W_{n,l}'(s).
\label{eq:NCrelation}
\end{eqnarray}    
Eq.\,(31) in the letter can be derived from these two equations.

%%%%%%%%%%%%%%%%%%%%%%%%%%%%%%%%%%%%%%%%%%%%%%%%%%%%%%%%%%%%%%%%%%%%%%%

%%%%%%%%%%%%%%%%%%%%%%%%%%%%%%%%%%%%%%%%%%%%%%%%%%%%%%%%%%%%%%%%%%%%%%%
%%%%%%%%%%%%%%%%%%%%%%%%%%%%%%%%%%%%%%%%%%%%%%%%%%%%%%%%%%%%%%%%%%%%%%%  

\end{document}